\documentclass[aps,prl,twocolumn,showpacs,superscriptaddress,groupedaddress]{revtex4}  
\usepackage{graphicx}  
\usepackage{dcolumn}   
\usepackage{bm}        
\usepackage{amssymb}   

\usepackage{wrapfig}
\usepackage{mhchem}
\usepackage{epstopdf} 

\begin{document}

\widetext


\title{Spatial Effects of Delay-induced Stochastic Oscillations in a Multi-Cellular System}
\author{Dmitry Bratsun} \affiliation{Department of Applied Physics, Perm National Research Polytechnical University, 614990, Perm, Russia}
\author{Andrey Zakharov}
\affiliation{Department of Chemical Engineering, Technion - Israel Institute of Technology, 32000, Haifa, Israel}


\begin{abstract}
We explore the joint effect of the intrinsic noise and time delay on the spatial pattern formation
within a multi-scale mobile lattice model of the epithelium. The protein fluctuations are driven by 
transcription/translation processes in epithelial cells exchanging chemical and 
mechanical signals and are described by a single-gene auto-repressor 
model with constant delay. Both deterministic and stochastic descriptions 
are given. We found that time delay, noise and spatial signaling can result 
in the protein pattern formation even when deterministic description exhibits 
no patterns.
\end{abstract}


\maketitle

The small number of reactant molecules involved in gene regulation can lead 
to significant fluctuations in protein concentrations, 
and there have been numerous studies devoted to the influence of such noise 
at the regulatory level since pioneering works in the early 2000s \cite{elston1,elston2,elow}. 
For a good review of recent developments, 
see~\cite{wil,tsi}.

The transcription-translation processes are 
compound multistage reactions involving the sequential 
assembly of long molecules.~It can provoke a time lag in gene regulation processes.~Until 
delays are small compared with other significant time scales characterizing 
the genetic system, one can safely ignore them in simulations.~However, 
if the lags become longer than other processes, 
the system has to be considered as non-Markovian, and one should account for it 
in both deterministic and stochastic descriptions.~The joint effect of the intrinsic noise 
and time delay on the temporal behavior 
during gene regulation have been studied first in~\cite{bratsun1,bratsun2}.~We 
have suggested a generalization of the Gillespie algorithm~\cite{doob,gil} 
widely used to simulate statistically correct trajectories of the state 
of a chemical reaction network that accounts for delay~\cite{bratsun1}.
Based on this technique, we showed that quasi-regular fluctuations 
can arise in the stochastic system with delay even when its deterministic counterpart 
exhibits no oscillations~\cite{bratsun2}.~Since 
that papers, there have been a lot of works developing this line of research 
(see \cite{hig,pahle,tsi,bur1} for recent reviews). They mainly focus on further 
improving the algorithm and on studying the temporal dynamics of the different 
gene systems with delays.

Several years ago, Lemerle {\it et al.}~\cite{lem} have noticed that 
``space is the final frontier in stochastic simulations of biological systems''.~The problem 
is that the massive amount of spatio-temporal experimental data 
has been accumulating, but stochastic models of biochemical processes are focusing mostly 
on the temporal dynamics. If in the past years spatial stochastic simulations 
of Markovian processes have made considerable progress~\cite{tsi,lem,chen,bur, bur2}, 
the examples of studies of non-Markovian stochastic systems are very rare.~For instance, 
Marquez-Lago {\it et al.}~\cite{bur3} discussed
how the spatial displacement of molecules can be incorporated into purely 
temporal models through distributed delays.~Danino {\it et al.}~\cite{dan} have 
given the remarkable experimental data with the pattern formation of delay-induced rhythms 
in a population of {\it E. Coli}, but did not provide the stochastic modeling. 
The theoretical difficulties seems to be clear: the generalization 
of the Gillespie algorithm to the case of the spatial dynamics of time-delayed 
processes is still waiting for its author.

\begin{figure}[b]
\center{\includegraphics[scale=0.2]{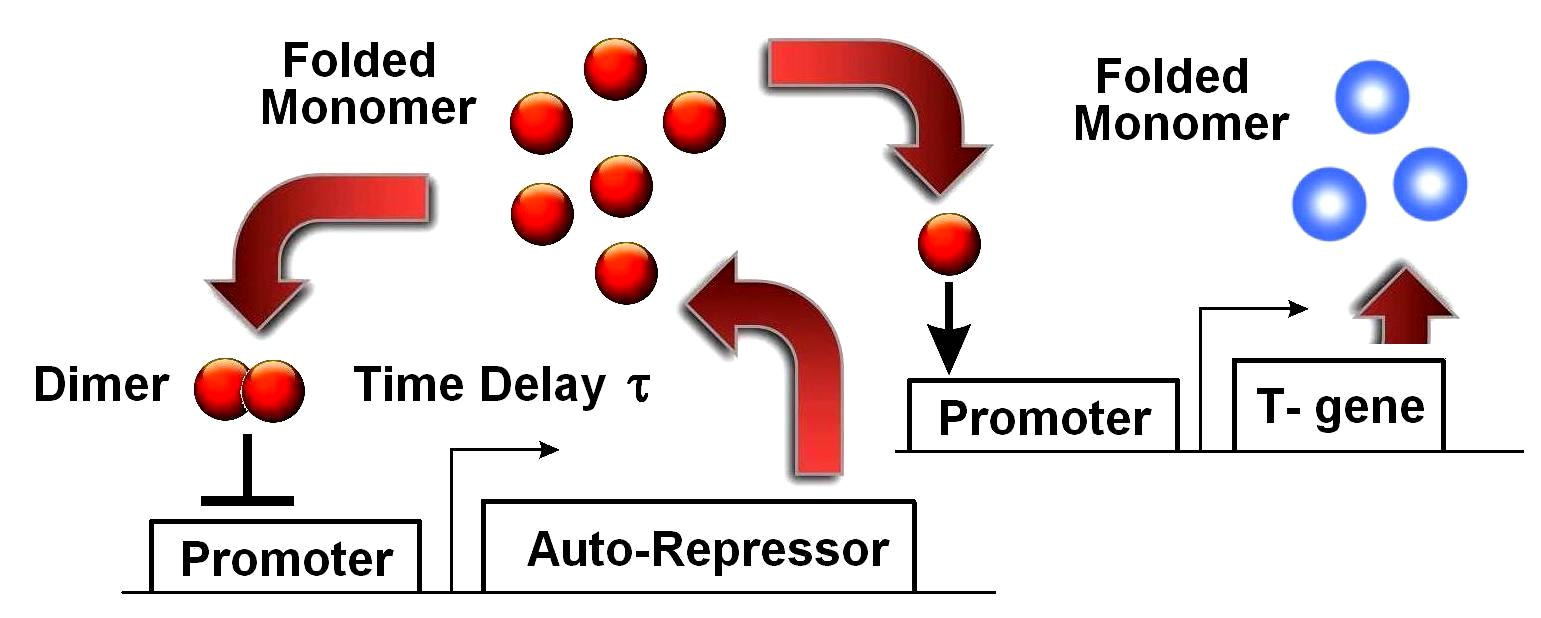}}\\
  \vspace{0.03in}
\center{\includegraphics[scale=0.2]{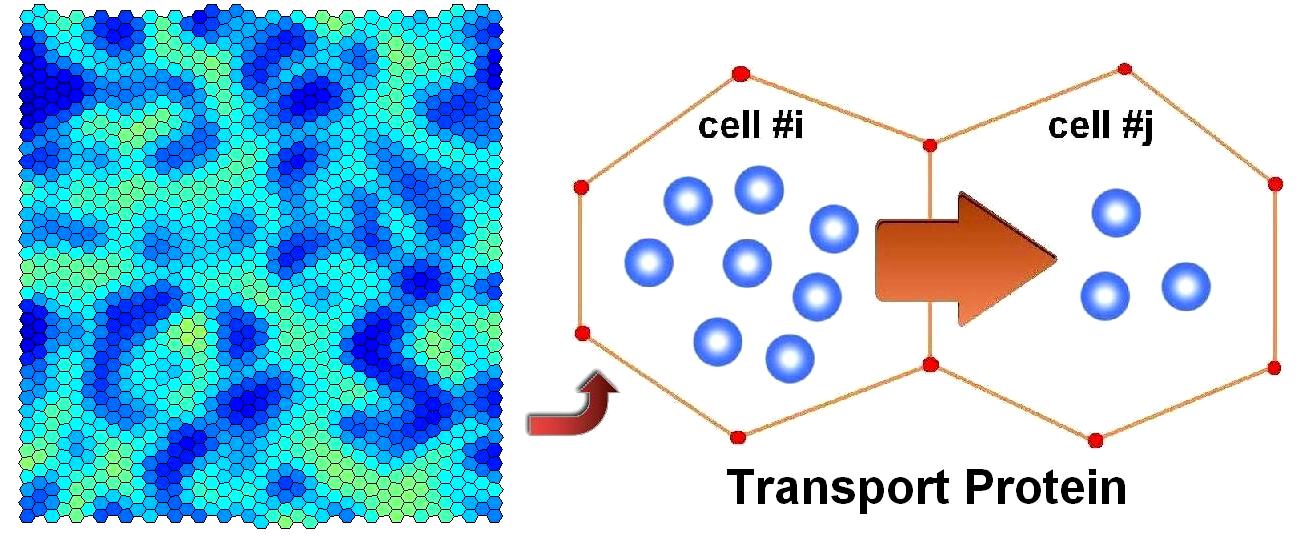}}
\caption{\label{fig:1} A single-gene autorepressor oscillator 
with a time-delayed negative feedback. 
Its protein (red circles) acts as a positive regulator of the transport protein (blue circles)
by activating its transcription. The signaling species is diffused from one cell 
to the other within the lattice governed by the chemo-mechanical 
epithelium model described in~\cite{pismen2,bratsun3}.}
\end{figure}

In this Letter, we explore the spatial effects produced by the combined action
of the intrinsic noise and time delay within a multi-cellular system. 
It is based on the multi-scale chemo-mechanical model of the epithelial 
tissue which first was developed in~\cite{pismen2} and then was applied 
to simulate the carcinoma growth~\cite{bratsun3}. 
Using the lattice approach allows to avoid the problem of the lack of a reliable 
Gillespie-like algorithm for spatial non-Markovian processes and to apply
the algorithm from~\cite{bratsun1}.
We use a single-gene auto-repressor model with constant delay at a single-cell level
complemented by the dynamics of the transport species penetrating cell membranes (Fig.~\ref{fig:1}).

Epithelial tissue is a two-dimensional layer of cells covering the surface of an organ or body.
The chemo-mechanical model includes the calculation of the dynamics 
of separate cells adapting to environmental stresses. 
The initial hexagonal lattice representing cells becomes distorted 
due to proliferation/intercalation and eventually incorporates also polygons 
with a different number of vertices. The epithelium is evolved by moving the cell nodes. 
The mechanical force acting on any $i$th node 
is defined via the elastic potential energy of the tissue~\cite{pismen2,bratsun3}:
\begin{equation}
{\bf F}_{i} = - \frac{1}{2} \frac{\partial}{\partial {\bf R}_i} \displaystyle\sum_{cells} 
\left(\kappa L^2 + \eta (S - S_0)^2\right) + {\bf F}^{st}_i ,
\label{eq1}
\end{equation}
where ${\bf R}_i$ is the radious vector of $i$th node, $L$ and $S$ stand for 
the perimeter and area of a cell respectively. ${\bf F}^{st}_i$ is an uncorrelated
zero mean stochastic input. The coefficient $\kappa$ defines the action 
of contractile forces within the cell cortex, $\eta$ reflects the cell resistance to any 
changes with respect to the reference area $S_0$. 

Since the motion is strongly overdamped, the governing equation for the displacement 
should be written as 
\begin{equation}
\frac{d{\bf R}_i}{d t} = {\bf F}_{i} H \left(|{\bf F}_{i}| - F_0\right) ,
\label{eq2}
\end{equation}
where $H$ is a Heaviside function, $F_0$ is the threshold 
force below which the node remains immobile. Altogether, Eqs.~(\ref{eq1}-\ref{eq2}) define 
the mechanics of the tissue. 

We consider a single-gene protein synthesis with negative auto-regulation (Fig.~\ref{fig:1}).~This 
is a popular motif in genetic regulatory circuits, and its temporal dynamics 
has been analyzed within both deterministic and stochastic framework~\cite{elston1}. 
This model is relatively simple yet but still maintains a high degree of biological relevance.
Its generalized version accounting for the effect of time delay $\tau$ 
has been suggested in~\cite{bratsun1,bratsun2}. Suppose that protein can exist both in the form 
of monomers X and dimers X$_D$. The transitions between them with the rates $k_{\pm d}$ are
\begin{equation}
\ce{ X + X ->[k_{+d}] X_D} , \qquad \ce{X_D ->[k_{-d}] X + X} .
\label{eq3}
\end{equation}
We assume that the protein may be degraded with the rate $B$ and be produced with the rate $A$ respectively:
\begin{equation}
\ce{X ->[B] \emptyset}, \qquad \ce{D_0^t ->[A] D_0^{t+\tau} + X^{t+\tau}}  .
\label{eq4}
\end{equation}
The synthesis occurs at time $t+\tau$ if the chemical state of the promoter site 
of the X gene at time $t$ is unoccupied (D$_0$). Otherwise (D$_1$), the production is blocked.~The 
transitions between the states occur with rates $k_{\pm 1}$
during binding and unbinding of some dimer respectively
\begin{equation}
\ce{D_0 + X_D ->[k_{+1}] D_1} , \qquad  \ce{D_1 ->[k_{-1}] D_0 + X_D}  .
\label{eq5}
\end{equation}

In order to describe the intercellular signaling, 
we introduce the transport species T positively regulated 
by the X protein (Fig.~\ref{fig:1}). 
If the operator site of the T gene is occupied (D$_1^T$) then the T protein
may be produced immediately with a certain probability $A_T$:
\begin{equation}
\ce{{\emptyset} ->[A_{T}D^{T}_1(t)] T} .
\label{eq7}
\end{equation}
If the site is unoccupied, the production of the T protein is blocked.
Thus, the X monomers act as a positive regulator of T by activating its transcription,
because the transitions between the states with the rates $k_{\pm 2}$ are  
\begin{equation}
\ce{D_0^T + X ->[k_{+2}] D_1^T} , \qquad  \ce{D_1^T ->[k_{-2}] D_0^T + X}. 
\label{eq8}
\end{equation}
We assume also that once a signal has come in a certain cell,
it is converted into the X monomers with the rate $B_T$ and the copy number $N$:
\begin{equation}
\ce{T ->[B_T] X} .
\label{eq9}
\end{equation}
Altogether, Eqs.~(\ref{eq3}-\ref{eq9}) define the kinetics of gene regulation 
both at a single cell level and a whole tissue.

\begin{figure}
\center{\includegraphics[scale=0.33]{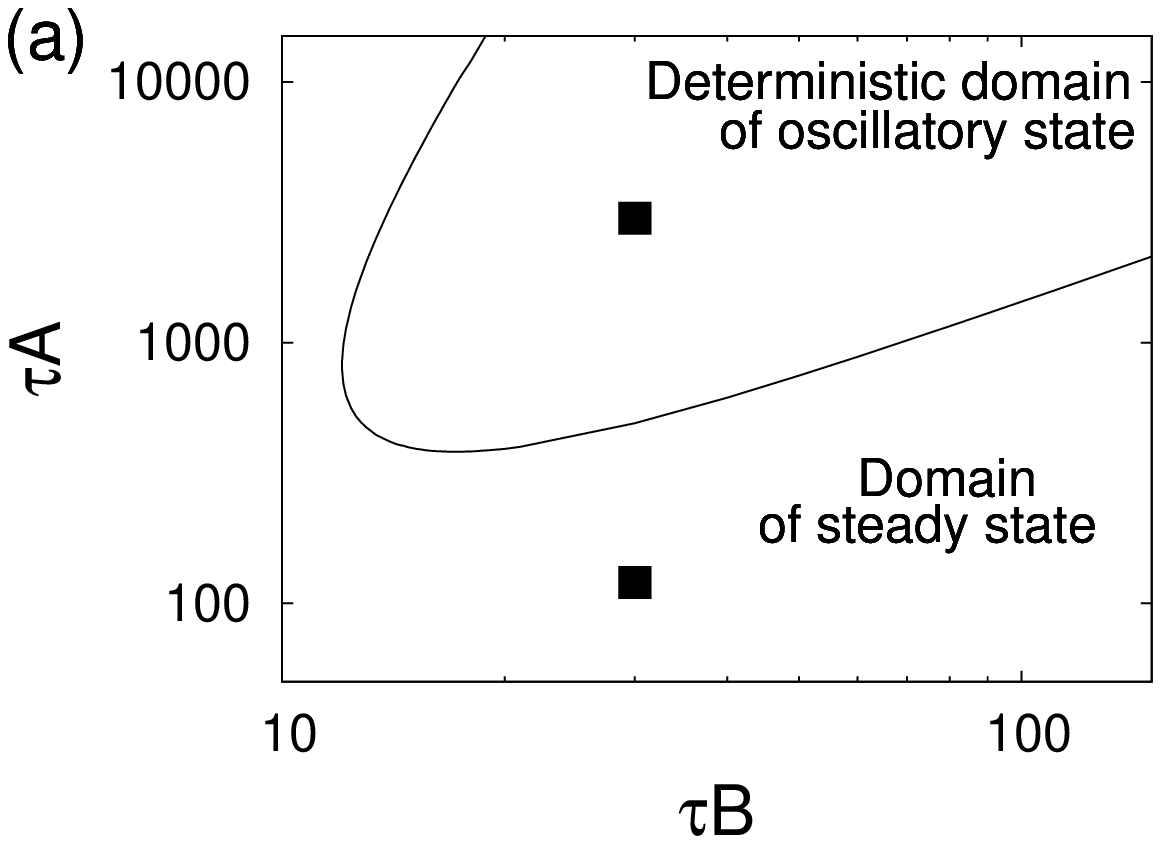} \includegraphics[scale=0.33]{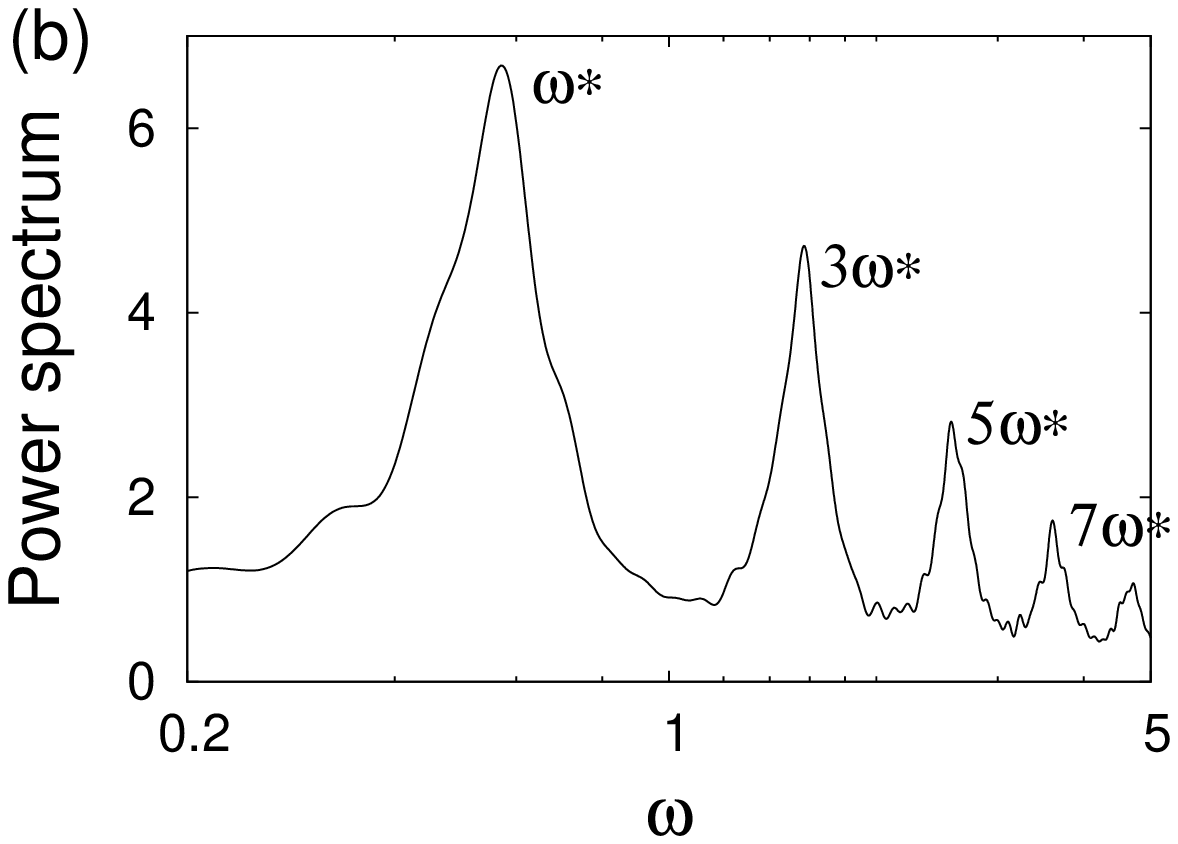}}
\caption{(a) Neutral curve of the Hopf bifurcation obtained within a single-cell deterministic description for $\varepsilon=0.1$, $\delta = 0.2$. (b) Power spectrum obtained in a single-cell stochastic simulation 
below the Hopf bifurcation for $A=20$, $B = 5$, $\tau = 6$, $k_{+1}=100$, $k_{+d}=200$, 
$k_{-1}=1000$, $k_{-d}=1000$.}
\label{fig:2}
\end{figure}

{\it Deterministic description}.~The main approximation is that 
the reactions of dimerization (\ref{eq3}) and binding/ unbinding (\ref{eq5},\ref{eq8}) 
are fast in comparison with production/ degradation 
of proteins (\ref{eq4},\ref{eq7},\ref{eq9}).~Thus, their dynamics has to enter quickly
into a local equilibrium and we arrive to
\begin{eqnarray}
(1 + 4 \varepsilon x_i){d x_i \over d t} = \frac{A}{1+\varepsilon\delta x_i^2(t-\tau)} + B_T \theta_i - B x_i , 
\label{eq10}
\\
\frac{d \theta_i }{d t} = \frac{N A_T \sigma x_i}{1+\sigma x_i} - B_T \theta_i +
\displaystyle\sum_{j\in adj(i)}  \alpha L_{ij}(\theta_j - \theta_i)  ,
\label{eq11}
\end{eqnarray}
where the subscripts refer to cells, $x$ and $\theta$ stand for the concentrations of 
the X and T proteins respectively, $\alpha$ is the transfer coefficient, 
$\delta= k_{+d}/k_{-d}$, $\varepsilon = k_{+1}/k_{-1}$, 
$\sigma=k_{+2}/k_{-2}$ and $adj(i)$ stands for ``adjacent to $i$-cell''.
It is assumed that the T protein is transported diffusively from one cell 
to the other, whereas its flux does not depend on the distance between the 
two cells $i$ and $j$ but is proportional to the boundary length $L_{ij}$.~This 
implies that the transport is limited by the transfer though 
cell membranes. The link between sub-cellular and macroscopic scales
is established through the Eq. (\ref{eq11}), since the $\theta$ field is global for 
the whole tissue (for more details, see Supplementary materials).

The neutral curve for the Hopf bifurcation derived within the deterministic 
approach (\ref{eq10}-\ref{eq11}) at a single-cell level is plotted in Fig.~\ref{fig:2}a 
in the plane of dimensionless parameters $\tau A$ and $\tau B$. The numerical study reveals
the common dynamics below and above the bifurcation.

\begin{figure}
\centering
 (a) \includegraphics[width=1.4in]{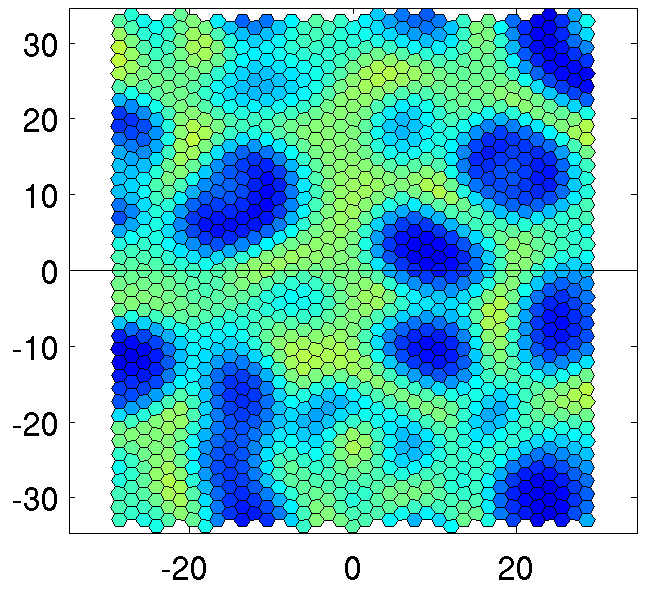}
  \includegraphics[width=1.74in]{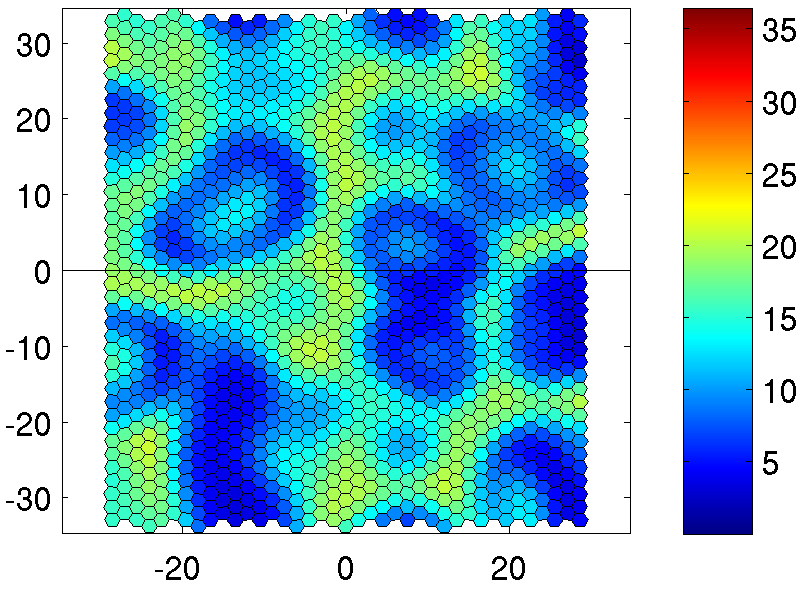}\\
 (b) \includegraphics[width=1.4in]{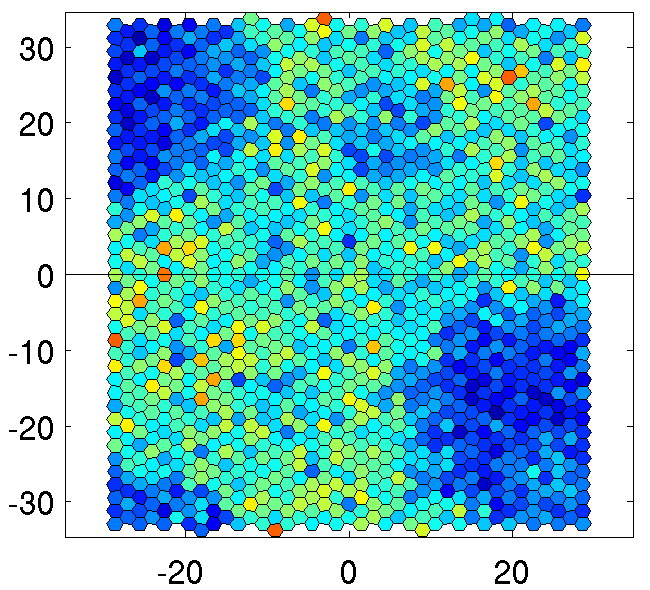}
  \includegraphics[width=1.74in]{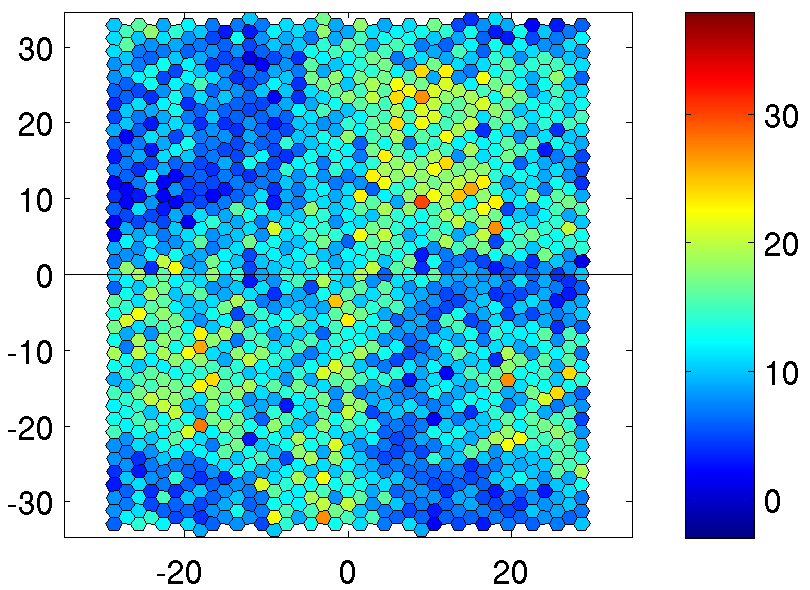}
  \caption{Evolution of the X protein pattern in the epithelium  
formed of more than 1500 cells within (a) deterministic and (b)
stochastic description, with parameters $A = 500$, $A_T=500$, $B = 5$, 
$\tau = 6$, $N = 1$, $\alpha=0.05$, $k_d =200$, $k_{-i}=1000$, $k_{1,2} =100$ 
taken above deterministic Hopf bifurcation curve (see Fig.~\ref{fig:2}a).
The frames correspond to times $t=260$, $340$.}
  \label{fig:3}
\end{figure}

In order to study the spatial effects, the set of delay differential equations (\ref{eq10}-\ref{eq11}) 
has been solved using the explicit Euler method, whose stability was warranted by a sufficiently 
small time step. This procedure was synchronized with the simulation of the mechanical 
evolution governed by Eqs. (\ref{eq1}-\ref{eq2}).~The initial configuration of the system 
is a hexagonal lattice comprising $1560$ cells with random phase distribution. 
The tissue as a whole has the form of a stripe. 
The typical values of the parameters governing the tissue mechanics 
hereinafter are as follows: $\kappa = 1.0$, $\eta = 1.0$, $S_0 = 3\sqrt{3}/2$,  
$F_0 = 0.02$. Fig.~\ref{fig:3}a presents the results of numerical simulation 
with parameters taken above the Hopf bifurcation curve (the upper black square in Fig.~\ref{fig:2}a). 
The X protein patterns are shown for two consecutive moments of time. 
The nonlinear dynamics includes the slow development of spiral traveling wave 
pattern which arises against the synchronized field oscillating in the background.
The oscillation period is approximately equal to the triple delay time.

{\it Stochastic description}. In order to describe the spatial stochastic effects, 
we use a hybrid model, which is constructed as follows.
The dynamics of the proteins in each cell has been obtained by 
performing direct stochastic simulations of the reactions (\ref{eq3}-\ref{eq9})
using the modified version of the Gillespie algorithm~\cite{bratsun1}.
The signaling between cells is still organized as diffusive transport 
from one cell to the other according to finite-difference formula:
\begin{eqnarray}
T_i^{t+\triangle t} = T_i^{t} + \left[\triangle t \displaystyle\sum_{j\in adj(i)}  
\alpha L_{ij}(T_j^t - T_i^t)\right] , 
\label{eq12}
\end{eqnarray}
where $[...]$ stands for the integer part of the expression.
The time step in (\ref{eq12}) 
is equal to the time step for the integration of (\ref{eq1}-\ref{eq2}): 
$\triangle t = 0.05$. Since the typical time step stochastic system is much less ($\triangle t_{st}=0.00001-0.001$), one needs to dock the numerical schemes 
for the mechanical evolution of the tissue and stochastic fluctuations of the protein (see Supplementary materials).

\begin{figure*}
\centering
  \includegraphics[width=1.4in]{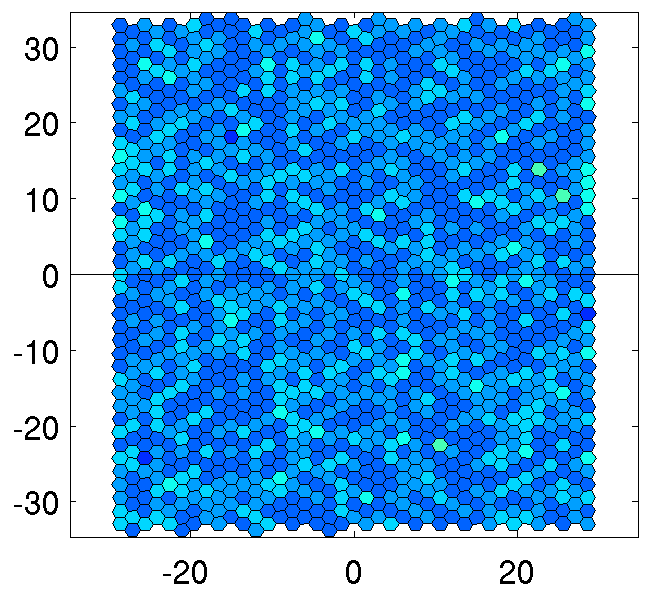}
  \includegraphics[width=1.4in]{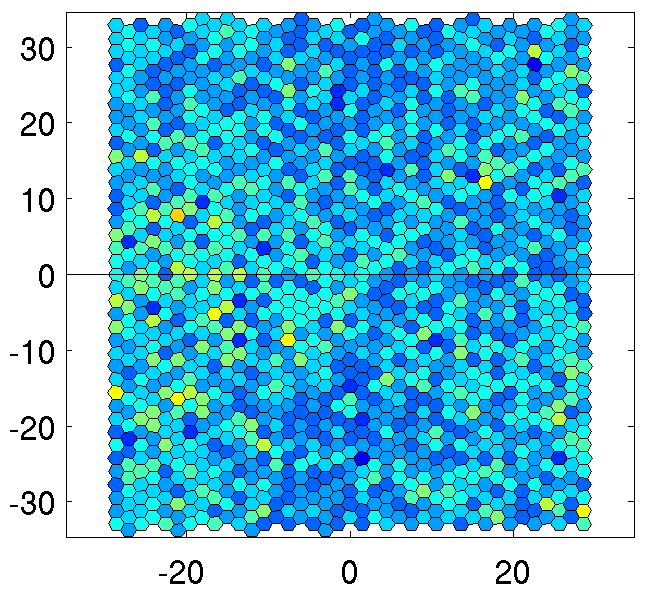}
  \includegraphics[width=1.4in]{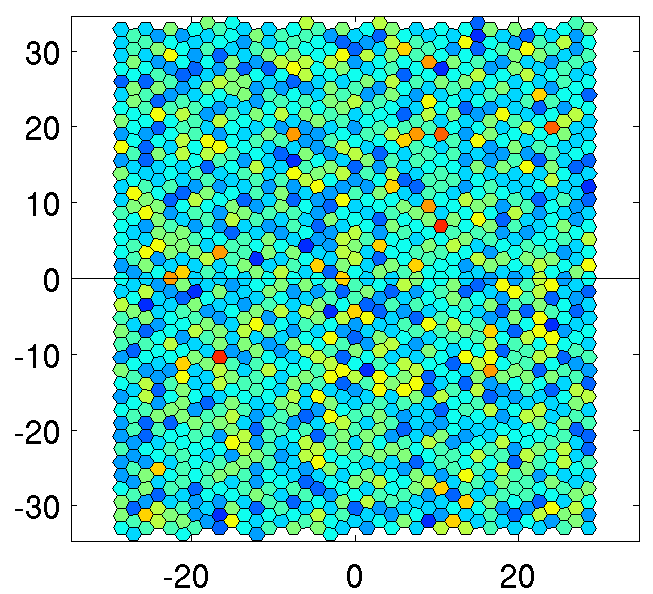}
  \includegraphics[width=1.75in]{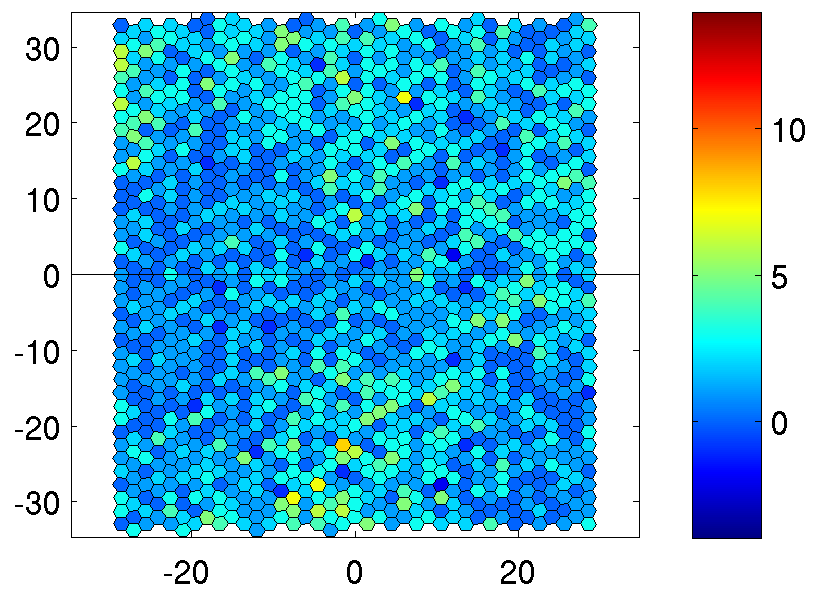}\\
  \caption{Stochastic pattern formation of the X protein in the epithelium far below 
the Hopf bifurcation (see Fig.~\ref{fig:2}a). The frames correspond 
to times within one oscillatory period $2\tau$. The parameters are the same 
as in Fig.~\ref{fig:3}, except $A = 20$, $N=8$.}
  \label{fig:4}
\end{figure*}

The power spectrum of time-delayed stochastic signal generated on a single-cell level 
far below the Hopf bifurcation is shown in Fig.~\ref{fig:2}b. 
It has a remarkable row of peaks indicating that some periodicity 
in the signal should exist. The first peak corresponds to the frequency 
$\omega^* = \pi /\tau \approx 0.524$ ($\tau=6$). One can observe also 
the strong harmonics of the fundamental frequency given by $\omega_n = (2n - 1) \omega^*$.
Thus, the time delay coupled with the noise can cause a system 
to be oscillatory even when its deterministic description predicts no oscillations.

Let us consider now two examples of spatial stochastic simulations. Fig.~\ref{fig:3}b 
presents the stochastic pattern formed by the X monomers in the tissue  
with the same parameter values as in Fig.~\ref{fig:3}a. 
We found that nonlinear dynamics of spatially extended system consists of
two distinct oscillatory modes, just like it was in the deterministic case. 
One is a quasi-standing wave pattern oscillating 
with $\omega^*$. The second oscillatory mode consists of traveling waves 
which arise from selected initial disturbances. In fact, the stochastic
pattern looks very similar to its deterministic counterpart obtained for the same
parameters (compare with the upper row in the same figure). 
The wavelength of the structure is found to depend on the copy 
number $N$ whose growth enhances fluctuations and diffusive 
fluxes between cells. 

\begin{figure}
\centering
  \includegraphics[width=2.9in]{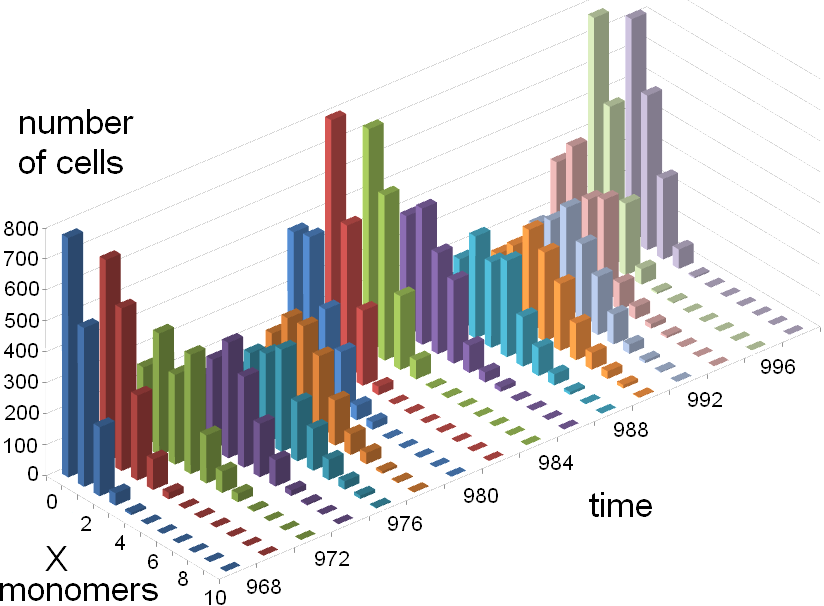}
  \caption{Time evolution of the distribution of stochastic oscillation phases in cells
obtained far below deterministic Hopf bifurcation.~Two periods of spatially synchronized oscillations
are shown.~The parameters are the same as in Fig.~\ref{fig:4}.}
  \label{fig:5}
\end{figure}

Consider now the parameters taken far below the Hopf bifurcation curve (Fig.~\ref{fig:2}a). 
We found that starting with random initial conditions, the system fairly quickly falls 
into a fully synchronized mode of oscillations 
with a common frequency $\omega\approx\omega^*$ (Fig.~\ref{fig:4},~\ref{fig:5}). 
We found also that the increase of the copy number $N$ can result in the clustering 
when the cells form two approximately equal communities, 
which oscillate in anti-phase. 
For instance, the numerical simulation with copy number $N = 4$ 
has showed that the clustering is not observed.
In contrast to that, this effect manifests itself 
clearly at $N =8$ after a sufficiently long integration (Fig.~\ref{fig:4}). In fact, the clustering in the system 
with a large amount of elements exchanging chemical signals has become 
at the center of attention of many scientists recently (see, for example,~\cite{kos}). 
It is believed that the clustering is likely to be the reason of further 
cells differentiation in organs. 

{\it Closing remarks}.~It is known that the noise during gene expression comes about 
in two ways.~The inherent stochasticity of transcription/translation generates {\it intrinsic} noise. 
The {\it extrinsic} noise refers to variation in identically-regulated quantities between different cells. 
In this paper, we have focused on spatial effects of intrinsic noise. 
A reliable Gillespie-like algorithm for the spatial non-Markovian systems still has 
not been developed, but perhaps, this algorithm is not very necessary.
Since any living matter consists of cells, it is more practical to use 
numerical methods based on a computational mesh which represents
a cellular compartment, such as a membrane or the interior of some part of a cell.
In this paper, we have applied such lattice approach suggesting 
the model which includes both mechanical interactions and chemical signal exchanges 
between cells. 

An important new result of this study is giving insight into how the excitation 
of quasi-regular delay-induced fluctuations found in~\cite{bratsun1}, manifests itself 
in space.~We show that above the Hopf bifurcation it is observed the traveling wave pattern which 
is similar to that in the deterministic case. But more interesting result is found below 
the Hopf bifurcation where the deterministic system is stable: 
there may be observed both a spatial synchronization of oscillations and 
clustering of cell community. 

We wish to thank L.M. Pismen for stimulating discussions. 
The work was supported by the Perm Ministry of Education 
and RFBR (grant 14-01-96022r\_ural\_a).

\end{document}